# Isolating and enhancing the emission of single erbium ions using a silicon nanophotonic cavity

A.M. Dibos[*], M. Raha[*], C.M. Phenicie[*], and J.D. Thompson[†]

*Department of Electrical Engineering, Princeton University, Princeton, NJ 08544, USA*

**The ability to distribute quantum entanglement over long distances is a vital ingredient for quantum technologies[1,2]. Single atoms and atom-like defects in solids are ideal quantum light sources and quantum memories to store entanglement[3-8]. However, a major obstacle to developing long-range quantum networks is the mismatch between typical atomic transition energies in the ultraviolet and visible spectrum, and the low-loss propagation band of optical fibers in the infrared, around 1.5 μm. A notable exception is the $Er^{3+}$ ion, whose 1.5 μm transition is exploited in fiber amplifiers that drive modern communications networks[9]. However, an optical interface to single $Er^{3+}$ ions has not yet been achieved because of the low photon emission rate, less than 100 Hz, that results from the electric dipole-forbidden nature of this transition[10]. Here, we demonstrate that the emission rate of single $Er^{3+}$ ions in a solid-state host can be enhanced by a factor of more than 300 using a small mode-volume, low-loss silicon nanophotonic cavity. This enhancement enables the fluorescence from single $Er^{3+}$ ions to be clearly observed for the first time. Tuning the excitation laser over a small frequency range allows dozens of distinct ions to be addressed, and the splitting of the lines in a magnetic field confirms that the optical transitions are coupled to the $Er^{3+}$ ions' spins. These results are a significant step towards long-distance quantum networks and deterministic quantum logic for photons, based on a scalable silicon nanophotonics architecture.**

Single atoms and atom-like defects have been used to demonstrate a variety of key tasks for quantum networks, including spin-photon entanglement[4-6], entanglement of remote atomic spins[7,8], and deterministic quantum gates between photons[11,12]. This work has been carried out in a variety of physical systems, ranging from laser-cooled atoms[7,11,12] and ions[4], to quantum dots[6] and atomic defects in diamond[5,8,13,14].

In these experiments, the long-distance distribution of entanglement is achieved by sending a photon emitted by the atom through an optical fiber. A central challenge to extending this work beyond the laboratory scale is that the losses in standard optical fibers are relatively large for photons at the wavelength of previously studied atoms and atomic defects, ranging from approximately 8 dB/km (for $NV^-$ centers at 637 nm[8]) to 2 dB/km (for $SiV^0$ and GaAs quantum dots at 950-980 nm[6,14]). In comparison, the losses in the 1.5 μm "telecom band" are only 0.2 dB/km[9], which results in a 90 dB improvement in transmission over a modest 50 km link (relative to 2 dB/km). This has motivated significant efforts to perform single-photon wavelength conversion to 1.5 μm using nonlinear optics[15]. An alternative approach is to avoid fiber altogether using free-space transmission via satellites[16].

---

[*] These authors contributed equally to this work.
[†] Email: jdthompson@princeton.edu

In this work, we pursue a direct solution to this challenge based on single $Er^{3+}$ ions in a solid-state host, which have an optical transition with a wavelength of 1.5 μm. In addition to offering lower losses in fibers, this wavelength also enables integration with technologically mature silicon nanophotonic devices[17], which we employ here. Like other rare earth ions, $Er^{3+}$ features coherent spin[18] and optical[19] transitions even in solid-state hosts, as the active 4$f$ electrons are situated close to the nucleus and therefore only weakly coupled to phonons in the host crystal[10]. These properties have motivated the development of quantum memories for light based on rare earth ion ensembles[20,21]. However, observations of single rare earth ions have been hampered by the electric dipole-forbidden nature of intra-4$f$ optical transitions, which results in long excited state lifetimes and correspondingly low photon emission rates[10]. Consequently, optical emission from single rare earth ions ($Pr^{3+}$ and $Ce^{3+}$) has only recently been observed[22-25], although single $Er^{3+}$ ions in silicon have also been detected using a charge-sensing approach[26].

The key idea of our experimental approach is to enhance the emission rate of single $Er^{3+}$ ions by positioning an Er-doped crystal in close proximity to a silicon photonic crystal (PhC) cavity tuned to the transition frequency of the ion (Fig. 1a)[27,28]. The enhancement, denoted by the Purcell factor $P$, is maximized for small mode-volume, low-loss cavities. Silicon PhCs capable of achieving $P > 10^5$ have been demonstrated[29,30], which would result in photon emission rates from single $Er^{3+}$ ions of more than 10 MHz, despite the low initial rate of $\Gamma_0 = 2\pi \times 14$ Hz[31].

Our devices consist of one-dimensional silicon PhCs on a lightly Er-doped yttrium orthosilicate ($Y_2SiO_5$, or YSO) substrate. We fabricate the silicon PhCs from a silicon-on-insulator wafer using electron beam lithography and reactive ion etching, then transfer them onto YSO using a stamping technique. YSO is chosen as a host because it is available in high-quality, transparent single crystals, and Er substitutes easily for $Y^{31}$; the Er concentration is approximately 3 ppm in our experiments. Ions near the YSO surface couple to the cavity through the evanescent electric field, whose magnitude $|E|$ at the Si-YSO interface is 60% of its maximal value in the center of the Si layer. The substrate is mounted on a cold finger inside a closed-cycle cryostat (T ≈ 4.2 K). A lensed fiber couples light to and from the single-sided cavity with around 50% one-way efficiency, and a fiber-coupled superconducting nanowire single photon detector (SNSPD) located in a second cryostat detects light leaving the cavity (Fig. 1c). We tune the frequency of the cavity resonance *in situ* by condensing gas on the surface of the device. Additional details about the fabrication and measurement techniques are discussed in the supplementary information (SI).

We search for ions coupled to the cavity using photoluminescence excitation spectroscopy (PLE), with the pulse sequence shown in Fig. 1d. We record a spectrum by scanning the laser frequency and cavity resonance together through a spectral region near the Er:YSO (site 1) bulk absorption resonance at 1536.46 nm. The resulting spectrum features a series of sharp peaks (Fig. 2a), which we interpret as the optical transitions of individual $Er^{3+}$ ions. The width of the individual peaks is approximately 5 MHz (Fig. 2b). The height of a single peak above the background saturates at high excitation powers to about 0.02 detected photons following each excitation pulse (Fig. 2b, inset). The combined detection and collection efficiency of light in the cavity is 0.04. Therefore, the observed saturated count rate is consistent with a single ion with nearly perfect emission into the cavity and an incoherent excitation probability of 0.5. The inhomogeneous distribution of the individual ions' transitions results from local variations in the crystal environment caused by strain and proximity to other defects. The width of this distribution, around 5 GHz, is slightly larger than the 1 GHz width observed in bulk YSO crystals with very low $Er^{3+}$ concentration[31]. However, it is substantially smaller than the width in glassy hosts such as silica fibers, which is typically of order 10 THz[10].

To quantify the strength of the atom-cavity coupling, we focus on a single peak in the inhomogeneous distribution and extract the excited-state lifetime τ from the time constant of the fluorescence decay (Fig. 3a). The measured value, τ = 45 ± 1 μs, is 252 ± 4 times shorter than the bulk lifetime of $τ_0$ = 11.4 ms for $Er^{3+}$:YSO. To confirm that the increased decay rate results from resonant enhancement by the cavity, we tune the cavity away from the atomic transition while keeping the laser frequency fixed, and observe that the decay rate, Γ = 1/τ, decreases to an asymptotic value consistent with $Γ_0 = 1/τ_0$ (Fig. 3b). From the maximum decay rate obtained when the cavity is resonant with the atomic transition, we conclude that the maximum Purcell factor for this ion is 320 ± 12. Using the relationship $P = 4g^2/(κΓ_0)$, we determine the complete cavity-QED parameters $(g, κ, Γ_0) = 2π × (2.08$ MHz, 3.85 GHz, 14 Hz) for this ion (here, $g$ is the single-photon Rabi frequency and κ is the cavity decay rate). The value of $g$ is roughly consistent with our theoretical prediction of $2π × 2.62$ MHz for an ion at the Si-YSO interface (see SI), where the discrepancy could be attributed to the position of the ion or a misalignment between the atomic dipole moment and the local cavity polarization.

To confirm that the sharp spectral peaks result from the transitions of individual $Er^{3+}$ ions, we measure the second-order autocorrelation function, $g^{(2)}$, of the fluorescence (Fig. 3c). The value of $g^{(2)}(0) = 0.36 ± 0.03 < 0.5$ indicates that more than half of the detected photons originate from a single emitter. It is also consistent with an estimate of accidental coincidences resulting from an independently measured background of dark counts and weakly coupled ions. The autocorrelation also shows bunching ($g^{(2)} > 1$) that decays on a timescale of 800 μs, whose origin is presently unknown.

Lastly, we apply a magnetic field using a permanent magnet and observe that a single-ion peak splits into two lines (Fig. 4). The magnetic field is oriented along an axis at an angle of 45 degrees to the YSO D1 axis, in the D1-D2 plane (Fig. 1a). The measured splitting is consistent with the difference of the ground and excited state g-factors (Δg = 1.55) for this orientation of the magnetic field[32]. The doublet nature of the ground and excited states (Fig. 1b) should give rise to four distinct transitions. However, bulk Er:YSO has been shown to have highly spin-conserving optical transitions[33], consistent with our observation of only two lines. Whether this behavior should persist in a cavity is unclear, since the branching ratio can change in a cavity with $P \gg 1$, because of polarization-dependent coupling to the cavity. Nevertheless, the brightest lines that we study likely result from ions where the stronger, spin-conserving dipole moment is aligned with the local cavity polarization, thus maintaining or even enhancing the spin-conserving nature. These measurements show that the optical transitions are coupled to the spin, which is crucial for future spin-photon entanglement.

We note that we have not been able to observe any dynamics of the ground state spin, such as optical spin polarization. This leads us to conclude that either the optical transitions are extremely spin conserving, that the spin-lattice relaxation time ($T_1$) is short compared to the optical excited state lifetime, or both. Prior measurements of $T_1$ for Er:YSO reveal a steep temperature dependence around 4 K, with $T_1$=1.5 ms at 4 K, but only 7.5 μs at 6 K[34]. Therefore, a slightly elevated sample temperature resulting from poor heatsinking in the cryostat could result in a significantly reduced $T_1$.

The experiments described above demonstrate that nanophotonic structures can be used to enhance and efficiently collect the photon emission from $Er^{3+}$ ions, enabling the observation of fluorescence from single $Er^{3+}$ ions for the first time. These results suggest several avenues for further investigation. First, operating at temperatures below 1 K should increase $T_1$ to more than $10^3$ seconds (see SI), to allow the exploration of spin-photon entanglement and coherent atom-photon interactions. While the longest spin coherence time $T_2$ observed for the ground state in Er:YSO is 6 μs[18] (believed to be limited by the $^{89}Y$ nuclear spin bath), longer spin coherence times may be achieved using fast dynamical decoupling, or by implanting $Er^{3+}$ ions into a host crystal without nuclear spins, such as silicon[35]. Second, implanting a small number of ions into an otherwise undoped YSO substrate will allow the number of ions coupled to the cavity to be controlled, and eliminate background fluorescence from distant, weakly-coupled ions. Finally, increasing the cavity quality factor to ten million[29] will increase the Purcell factor and lifetime-limited linewidth 200-fold; further improvement may be possible using smaller mode-volume cavity designs[30]. This will bring the Purcell-enhanced lifetime-limited linewidth close to the observed single-ion linewidth of 5 MHz, which is $10^3$ times broader than the current lifetime-limited linewidth $1/\tau = 2\pi \times 6$ kHz. The observed linewidth is presumably broadened by dephasing or fast spectral diffusion. Significantly lower spectral diffusion resulting in homogeneous linewidths of 74 Hz has been observed in bulk photon echo spectroscopy of Er:YSO at lower temperatures and in higher magnetic fields[19]. While the broadening mechanisms involved in low fields, as in the present work, are not well understood, future single-ion measurements will help to disentangle the contributions of phonons, ion-ion interactions, and coupling to the nuclear spin bath.

This work opens the door to realizing long-distance quantum networks based on a scalable and mature silicon nanophotonics platform. The ability to simultaneously couple the cavity to many spectrally resolvable atoms is promising for multiplexed repeater schemes, as envisioned with multimode ensemble quantum memories using rare earth ion ensembles[20]. The inhomogeneous broadening may be overcome by frequency-shifting photons during transit or using quantum eraser techniques, which are both possible because the total inhomogeneous linewidth is less than typical electronic bandwidths. Additionally, spectral addressing of ions that are spatially nearby (the average ion-ion separation is 32 nm in the present device) is a promising starting point for exploiting their electric or magnetic dipolar interactions for quantum logic[36] or quantum simulations of strongly interacting spin systems.

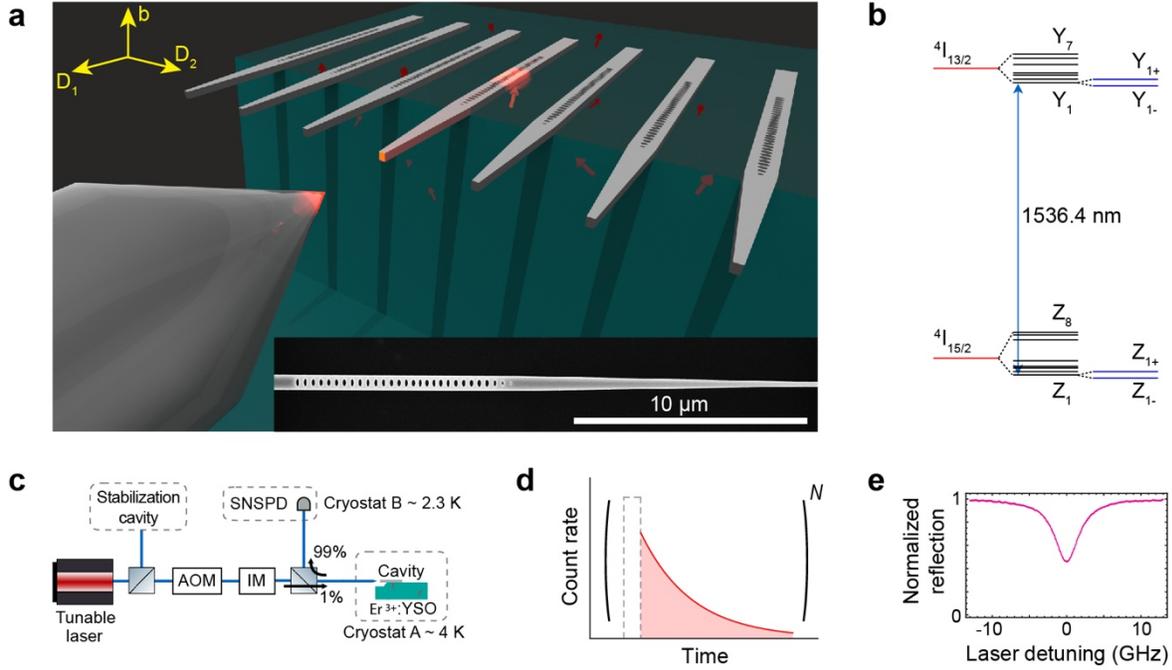

**Figure 1 | Experimental configuration for enhancing Er$^{3+}$ emission with a silicon photonic crystal. a,** Schematic illustration of the fabricated devices. Silicon waveguides patterned with photonic crystal cavities evanescently couple to Er$^{3+}$ ion impurities in a YSO crystal. The suspended, tapered ends of the Si waveguides protrude off the edge of the YSO crystal, allowing coupling to a lensed fiber. The axes (D1, D2, b) denote the orientation of the YSO crystal[32]. Inset: Scanning electron microscope image of a photonic crystal cavity and tapered waveguide prior to transfer onto the YSO substrate. **b,** Energy levels of Er$^{3+}$ in YSO. The crystal field splits the $^{2S+1}L_J$ states of the free Er$^{3+}$ ion (red lines) into J+1/2 Kramers' doublets (black lines), which further split into single states in a magnetic field (blue lines). The cavity is resonant with the $Z_1$-$Y_1$ transition at 1536.46 nm (YSO site 1). **c,** Schematic layout of the experiment. The Si-YSO device is situated in a cryostat (T ≈ 4.2 K). A stabilized laser in combination with a double-pass acousto-optic modulator (AOM) and electro-optic intensity modulator (IM) produces short pulses of light with an on/off ratio of 90 dB. A superconducting nanowire single photon detector (SNSPD) inside a second cryostat detects the return light from the cavity. **d,** The PLE measurement sequence consists of a 10 μs excitation pulse, followed by a fluorescence collection window. **e,** Reflection spectrum of the cavity used in these experiments, with quality factor $Q = 5.1 \times 10^4$.

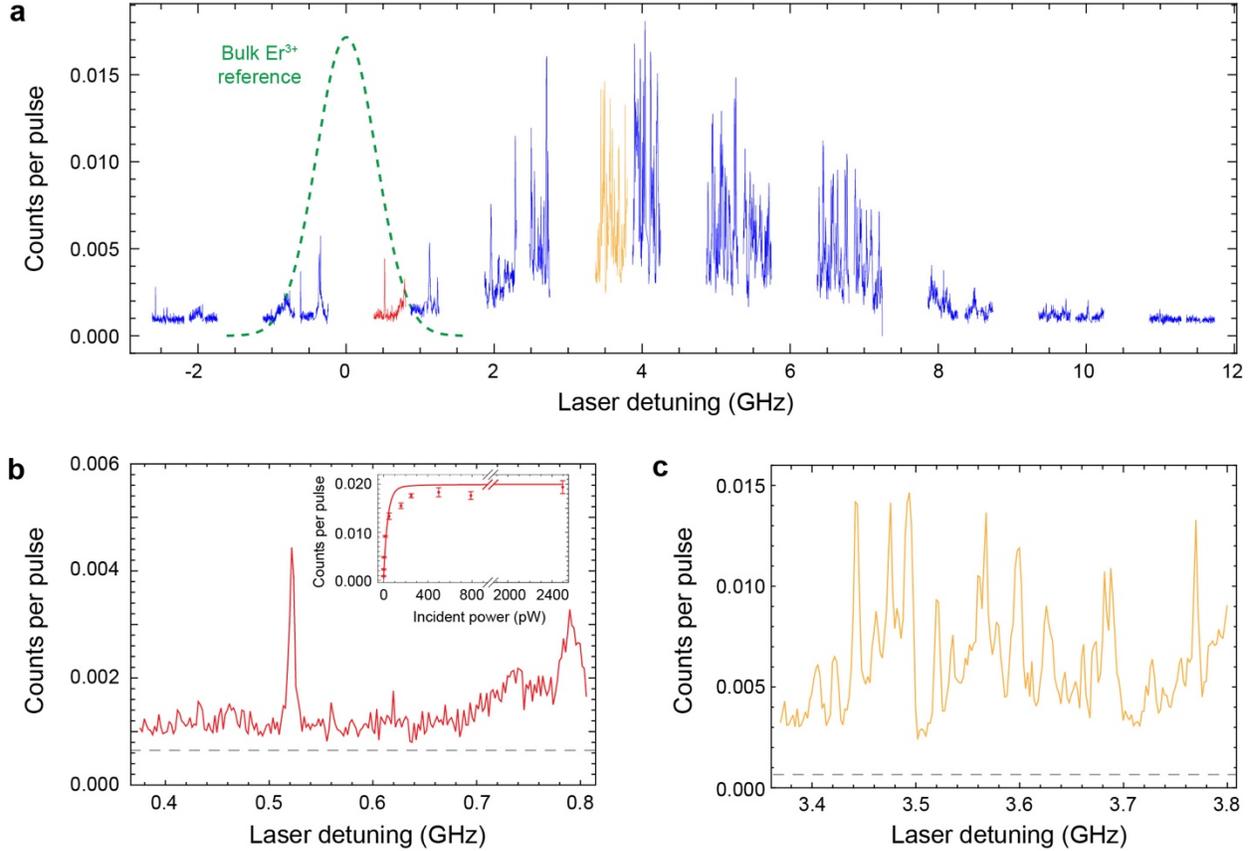

**Figure 2 | Photoluminescence excitation (PLE) spectrum of single ions in a dilute $Er^{3+}$ ensemble. a,** PLE spectrum, measured by scanning the laser frequency and cavity resonance (with linewidth $\kappa = 2\pi \times 3.85$ GHz) together through a spectral region near the bulk Er:YSO absorption resonance (determined from a second crystal, and indicated schematically by the green dashed line). The vertical axis indicates the average number of photons detected in an 82 μs integration window following each excitation pulse (as in Fig. 1d). The spectrum shows individually resolvable peaks, which we interpret as the optical transitions of single $Er^{3+}$ ions. The interruptions in the scan result from spectral regions that are inaccessible with our laser stabilization technique. **b,** Expanded view of the red portion of (a), showing an isolated line with a width of 5 MHz (full-width at half-maximum, FWHM) on a nearly dark-count-limited background (gray dashed line), characteristic of the tails of the inhomogeneous distribution. Inset: The peak height above the background saturates with increasing excitation power. The solid line is a model based on independently measured parameters (see SI). **c,** Expanded view of the yellow region in panel (a) near the center of the inhomogeneous distribution, where the spectral lines overlap.

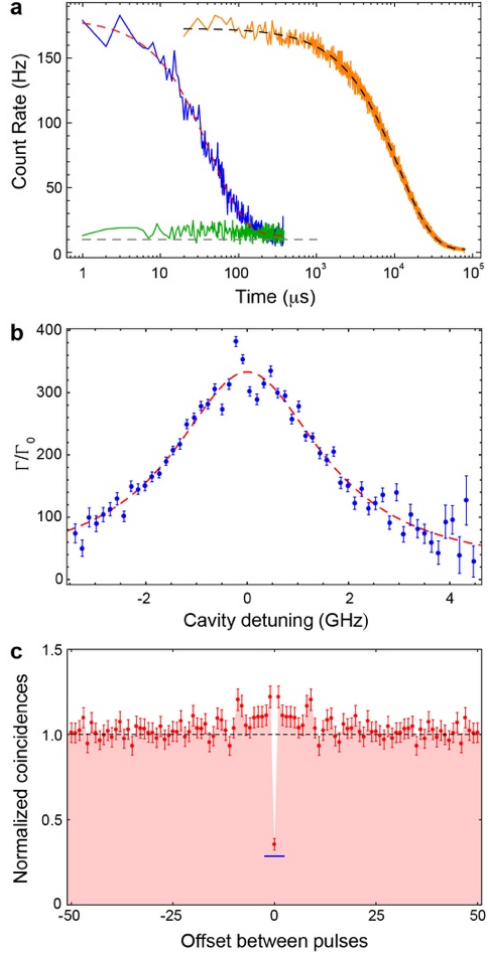

**Figure 3 | Quantifying the $Er^{3+}$-cavity coupling, and measurement of $g^{(2)}$. a,** Time-resolved fluorescence from a single cavity-coupled $Er^{3+}$ ion (blue) following an excitation pulse. An exponential fit (red dashed line) gives the excited-state lifetime $\tau = 45.3 \pm 0.7$ µs, which is $252 \pm 4$ times shorter than the bulk excited state lifetime of $\tau_0 = 11.4$ ms (orange; measured in a second crystal, and rescaled on the vertical axis). Repeating the measurement with the laser detuned by 35 MHz from the fluorescence peak shows that some background fluorescence is present (green) above dark counts (gray dashed line), but the absence of a fast time decay suggests that it originates from ions that are only weakly coupled to the cavity. **b,** Fixing the laser frequency to the atomic transition and sweeping the cavity resonance reveals that the decay rate enhancement $\Gamma/\Gamma_0$ varies with the atom-cavity detuning. A Lorentzian fit to the data (red dashed curve) yields a width of $3.6 \pm 0.3$ GHz (consistent with the cavity linewidth), a maximum decay rate enhancement $P = \Gamma/\Gamma_0 = 320 \pm 12$, and an asymptotic offset of $13 \pm 13$. **c,** Second-order autocorrelation function ($g^{(2)}$) of the fluorescence from a single ion. We bin all photons detected after a single excitation pulse into a single time bin, so the horizontal axis shows the autocorrelation offset in units of the pulse repetition period (100 µs). The data is symmetric around zero offset since a single detector is used to record the fluorescence and compute a true autocorrelation; however, both positive and negative offsets are plotted for clarity. The normalized number of coincidences at zero offset is $g^{(2)}(0) = 0.36 \pm 0.03 < 0.5$, which confirms that most of the photons detected originate from a single $Er^{3+}$ ion. This value is consistent with the value of $0.29 \pm 0.06$ (blue line) expected given the signal-to-background ratio of $5.5 \pm 1.6$ in this experiment (see SI).

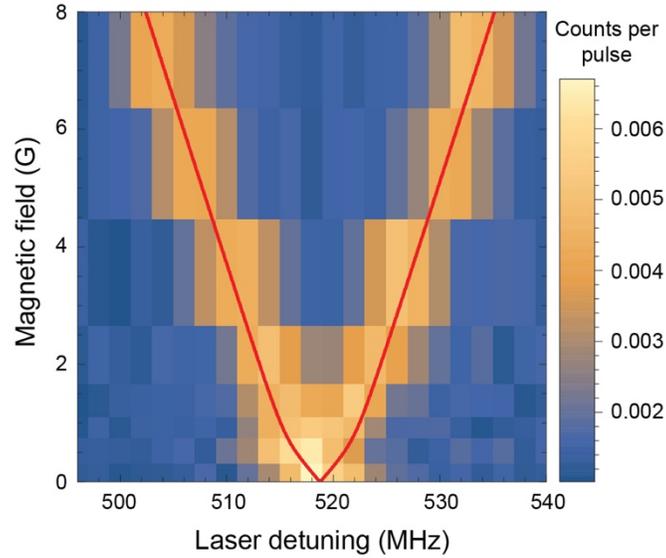

**Figure 4 | Zeeman splitting of a single Er$^{3+}$ ion spectral line.** A magnetic field splits a single spectral line into two transitions. The field is oriented approximately 45 degrees between the D1 and D2 axes in the D1-D2 plane. The observed splitting is consistent with the difference in the ground and excited state g-factors for this orientation of the magnetic field, with the inclusion of an offset field of around 1 G (red lines). Since the g-factors are highly anisotropic, the slope of the predicted splitting is not linear when the applied field is smaller in magnitude than the offset field, which has a different orientation.

**Acknowledgements**: We gratefully acknowledge helpful conversations with Nathalie de Leon, Andrei Faraon, Jevon Longdell and Mikael Afzelius, as well as technical assistance from Bert Harrop. Support for this research was provided by the NSF under the EFRI ACQUIRE program (grant 1640959). C.M.P. is supported by a NDSEG fellowship.

# Supplementary Information for
# Isolating and enhancing the emission of single erbium ions using a silicon nanophotonic cavity

A. M. Dibos*, M. Raha*, C. M. Phenicie*, and J. D. Thompson†

*Department of Electrical Engineering, Princeton University, Princeton, NJ 08544, USA*

## 1  Photonic crystal cavity design and theoretical Purcell factor

The photonic crystal cavity (PhC) used in our experiments is formed by a one-dimensional array of holes in a silicon waveguide, with an adiabatic reduction of the lattice constant in the center of the cavity to form a defect [1]. We developed the cavity design via simulation using the freely-available software packages MPB [2] and MEEP [3]. The YSO substrate is included in the simulations, but its slight birefringence is ignored, assuming an isotropic refractive index of $n = 1.80$. The principal field components of the cavity mode are shown in Fig. S1. With reference to the labels in Fig. S1a, the cavity dimensions are $(w_y, h_y, h_x, a_{cav}, a_{mir}) = (650, 315, 143, 295, 340)$ nm, where $a_{cav}$ is the lattice constant in the center of the defect and $a_{mir}$ is the lattice constant in the mirror region. The device thickness $w_z$ is 250 nm.

The key figure of merit is the electric field amplitude $|\vec{E}|$ in the YSO substrate, where the $Er^{3+}$ ions reside. Fig. S1c shows the variation of $|\vec{E}|^2$ across a cross-section of the device. The value of $|\vec{E}|^2$ at the Si-YSO interface is roughly 36% of its maximum value in the center of the Si waveguide, indicating that ions at the surface can still be well-coupled to the cavity even though they reside outside of the PhC itself. $|\vec{E}|^2$ decreases exponentially into the YSO substrate, falling by half every 45 nm.

The theoretical Purcell factor $P = \Gamma/\Gamma_0 - 1 = 4g(\vec{r})^2/\kappa\Gamma_0$ varies with the position of the ion, depending on the local single-photon Rabi frequency $g(\vec{r}) = d|\vec{E}(\vec{r})|/\hbar$, where $d$ is the electric dipole moment for the $Y_1$-$Z_1$ optical transition in Er:YSO, and $\vec{E}(\vec{r})$ is the electric field of a single photon in the cavity, at position $\vec{r}$. We determine the electric dipole moment $d$ from the excited state lifetime and previously measured branching ratio ($\beta = 0.21$) of the $Y_1 \to Z_1$ decay [4] using the following expression, including a local field correction [5]:

$$\Gamma_0 = \frac{1}{\beta} \cdot \left(\frac{3n_{\text{YSO}}^2}{2n_{\text{YSO}}^2 + 1}\right)^2 n_{\text{YSO}} \cdot \frac{d^2 \omega^3}{3\pi \epsilon_0 \hbar c^3}$$

where the spontaneous emission rate is $\Gamma_0 = 2\pi \times 14$ Hz, the refractive index is $n_{\text{YSO}} = 1.80$, and the frequency is $\omega = 2\pi \times 195$ THz. This gives $d = 2.8 \times 10^{-32}$ C-m, in general agreement with the value of $2.07 \times 10^{-32}$ C-m reported in Ref. [4]. The Purcell factor for different ion locations in the YSO is shown in Fig. S1d. The maximum value at the Si-YSO interface is 510, corresponding to $g = 2\pi \times 2.62$ MHz. The experimental value of $g$ for any particular ion will likely be lower than this because of sub-optimal positioning and imperfect alignment of the local cavity polarization with the (unknown) dipole moment orientation.

As discussed in the main text, intra-$4f$ electronic transitions are not electric dipole-allowed, as the various $4f$ states all have the same parity. In a host crystal, on a site without inversion symmetry (such as the Y site in YSO), the ligand field may mix $4f$ and $5d$ orbitals to introduce a weak "forced" electric dipole [6]. However, in the particular case of the $Er^{3+}$ $^4I_{13/2} \to {}^4I_{15/2}$ transition, the strength of the allowed magnetic dipole transition is comparable to that of the forced electric dipole [7]. While the above calculation of the Purcell factor assumed that the transition is purely electric, we note that the magnetic mode volume of the cavity is also small (Fig. S1b). Repeating the Purcell factor calculation while assuming a purely magnetic transition results in similar predicted enhancement ($g_{\mathrm{mag}} = 2\pi \times 4.39$ MHz and $P_{\mathrm{mag}} = 1.4 \times 10^3$). Designing structures to selectively enhance the electric or magnetic fields may allow the role of these transitions to be disentangled. Furthermore, interference between electric and magnetic dipole coupling to the cavity may be used to tailor the spin selection rules.

---

*These authors contributed equally to this work
†Email: jdthompson@princeton.edu

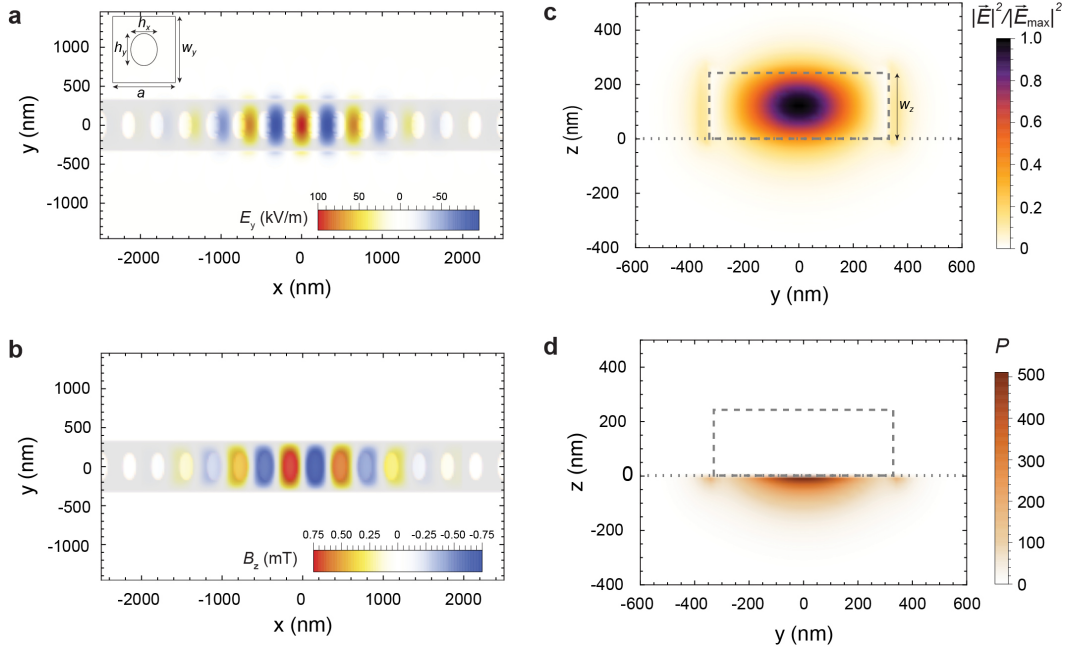

**Figure S1 | FDTD mode profiles and theoretical Purcell factors. a-b,** Top-view of the $E_y$ ($B_z$) component of the electric (magnetic) field in the photonic crystal cavity, in a cut through the middle of the Si layer (at $z = 125$ nm). **c,** Electric field $|E|^2$ in the $yz$-plane (at $x = 0$). The dashed rectangle indicates the Si waveguide; the region below the dashed line is the YSO substrate. $|E|^2$ is scaled with respect to its maximum value at the center of the waveguide, $|E_{\max}|^2$. **d,** Variation of Purcell factor as a function of position in the YSO substrate ($yz$-plane at $x = 0$), assuming purely electric dipole coupling to the cavity.

## 2 Fabrication process and materials

### 2.1 Fabrication process

The device fabrication consists of three distinct steps: (1) fabrication of Si PhCs, (2) preparation of the YSO substrate, and (3) transfer of the PhCs to the YSO.

We fabricate suspended Si PhCs from silicon-on-insulator (SOI) wafers using electron-beam lithography and plasma etching with $SF_6/C_4F_8/O_2$ gases in an inductively-coupled reactive ion etcher (SAMCO 800), followed by a hydrofluoric acid undercut (Fig. S2a). The SOI device layer is 250 nm thick, with a 2 $\mu$m buried oxide.

In order to couple light from the Si waveguides into a lensed optical fiber, the waveguides must protrude off the edge of the YSO substrate. The edges of the polished substrate are rounded and poorly defined. Therefore, we cut a sharp edge using a dicing saw, along with a progressively deeper series of relief cuts to give enough clearance for the lensed fiber.

After the undercut, we transfer the suspended Si devices onto the YSO substrate using a stamping procedure adapted from the stacking of 2D van der Waals materials [8](Fig. S2b). The process is carried out in a wafer bonder (Tresky T-3000-FC3), which controls the pressure, heating and alignment. In the waveguide design, we engineer pinch points which allow the suspended waveguides to be deterministically broken free from the SOI with the application of pressure using a PDMS stamp. The PhCs can then be pulled off the SOI wafer using a polypropylene carbonate (PPC) film on the stamp that is heated to 60°C to promote adhesion. Then, the stamp is positioned so a column of PhCs is aligned with the YSO edge, and pressed down onto the YSO. The stamp and YSO are heated to 130°C, which melts the PPC layer to release the PhCs and PPC from the PDMS stamp. Finally, we dissolve the PPC in chloroform, which leaves the PhCs in place on the YSO substrate (Fig. S2c). We align the PhCs on the substrate such that the principal electric field polarization ($E_y$ in Fig. S1) is parallel to the D2 axis. Previous spectroscopy has shown that site 1 Er:YSO ions have a larger absorption cross section for light with this polarization [9].

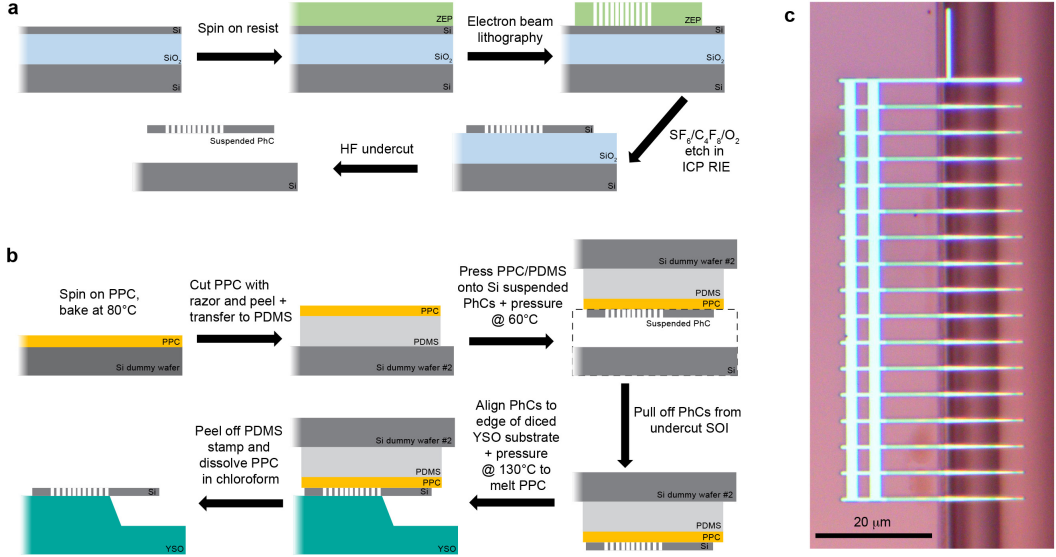

**Figure S2 | Fabrication processes for making Si-on-YSO cavities. a,** Side view diagram of the process flow for fabricating suspended Si photonic crystal cavities from an SOI wafer. **b,** The process flow for transfer of these Si photonic crystal cavities onto the diced edge of an erbium-doped YSO crystal. **c,** Top-down optical image of an array of PhCs stamped along the edge of the YSO substrate. The orientation of the photonic crystal cavities is the same as in panel (b). The vertical beam at the top of the image is used for alignment.

## 2.2 Substrate properties

The YSO crystal used in this work was grown, cut and polished by Scientific Materials (boule #07-0552). The RMS roughness of the surface is less than 0.5 nm (as measured by AFM), with occasional long scratches of 1-2 nm depth. The crystal has dimensions of $10 \times 10 \times 3$ mm, with the crystallographic orientation shown in Fig. 1a. It was not intentionally doped with erbium during growth, but trace quantities are present. Secondary ion mass spectrometry (SIMS) was used to estimate the concentration of Er (and other lanthanides), with the results summarized in Table 1. Because of the lack of a suitable reference with lower lanthanide concentrations, the measured quantities should be interpreted as upper bounds.

| Element | Ce | Pr | Nd | Sm | Gd | Dy | **Er** | Yb |
|---|---|---|---|---|---|---|---|---|
| Concentration (ppm) | 0.6 | 0.6 | 4.5 | 13.2 | 4.9 | 4.6 | **3.0** | 2.3 |

**Table 1 | Lanthanide concentrations from SIMS.** Upper bounds on the concentrations of several lanthanide elements from a 1.1 $\mu$m deep SIMS analysis of a YSO crystal from the same boule as the sample used in the experiments. The quoted concentrations are relative to the concentration of yttrium.

The number of ions that should be coupled to the cavity depends on the ion density in the substrate as well as the size and shape of the cavity mode. Using the simulation results in Fig. S1d, we can estimate the average number of ions that should experience a certain magnitude of Purcell enhancement (Fig. S3).

This may be compared to the experimental measurements shown in Fig. 2a of the main text (with additional analysis in Fig. S8). In the data, we estimate that at least several hundred single-ion peaks are visible. We believe that most of the peaks correspond to ions with Purcell factors greater than 50, given that the data was collected with a fluorescence integration window of 82 $\mu$s, which is much shorter than the bulk lifetime. This is consistent with an $Er^{3+}$ density between 0.3 and 3 ppm, in rough agreement with the SIMS measurement.

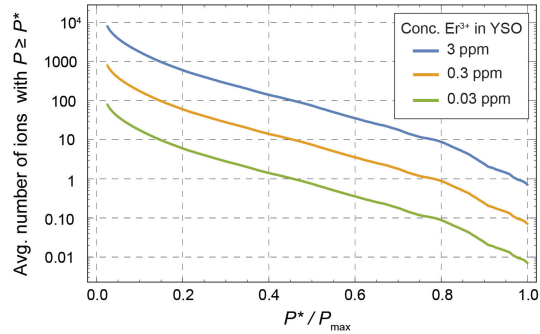

**Figure S3 | Number of ions with a given Purcell factor.** Average number of ions with a Purcell factor $P \geq P^{\star}$. $P^{\star}$ is normalized to the maximum Purell factor at the Si-YSO interface. In this figure, only ions at crystallographic site 1 are considered, since only these ions couple to the cavity resonance at 1536.4 nm.

## 3 Measurement and analysis techniques

### 3.1 Experimental configuration

In this section, we present a more detailed description of our measurement configuration (Fig. S4). The laser (Toptica CTL1500) frequency is coarsely stabilized by probing the absorption of a bulk Er:YSO crystal (50 ppm Er concentration) housed in the same cryostat as the PhC. Fine stabilization and linewidth narrowing are achieved using a stable reference cavity (Stable Laser Systems) housed in an ultrahigh vacuum chamber. To be able to tune the laser frequency while stabilizing to the reference cavity, we lock a sideband of the laser to the cavity. To achieve this, a small portion of the laser output passes through a phase modulating fiber EOM (PM) to add frequency sidebands at $\omega_{\text{laser}} \pm \omega_{\text{sb}}$. The sideband at $\omega_{\text{laser}} - \omega_{\text{sb}}$ is locked to the external cavity via the Pound-Drever-Hall technique, using sidebands applied by a second, free-space EOM, at 17.2 MHz. After locking, the sideband frequency is fixed to the cavity ($\omega_{\text{laser}} - \omega_{\text{sb}} = \omega_{\text{cav}}$), but the laser frequency itself can be tuned over a large range by changing $\omega_{\text{sb}}$. The stabilized laser linewidth is considerably less than 30 kHz. The light directed to the experiment is a pure tone without any locking sidebands.

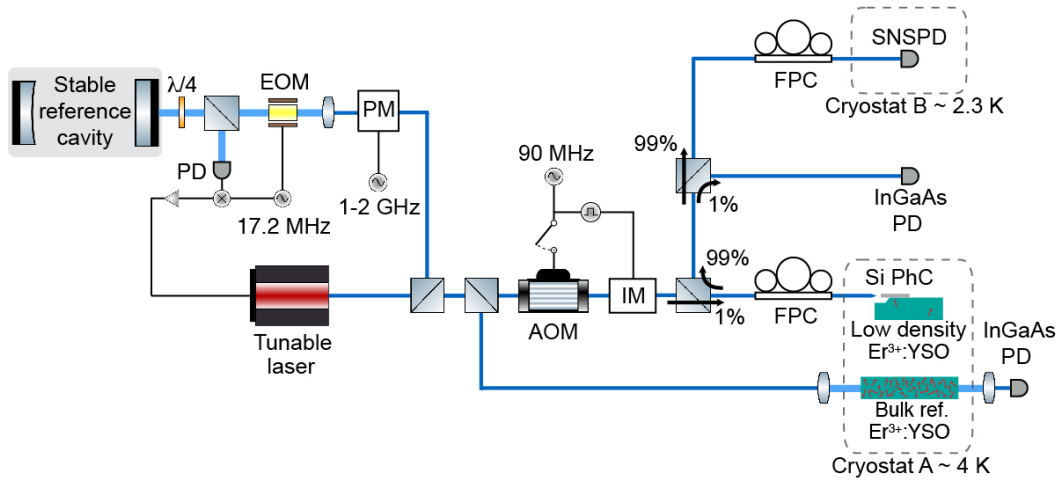

**Figure S4 | Experimental configuration.** A more detailed version of the experimental configuration than is presented in Fig. 1c of the main text. Abbreviations are explained in Sec. 3.1.

Laser pulses to excite the ions are produced using a double-pass acousto-optic modulator and intensity modulating electro-optic modulator in series (AOM and IM, respectively). This pair of modulators gives an on/off ratio of 90 dB. A fiber polarization controller (FPC) matches the laser polarization to the PhC cavity. The light returning from the cavity is detected by a superconducting nanowire single photon detector (SNSPD, Quantum Opus) in a separate cryostat. The current bias of the SNSPD is supplied by an SRS SG345 function generator and 100 k$\Omega$ resistor, which allows the bias to be switched off during an excitation pulse, preventing the detector from latching. An additional polarization controller adjusts the polarization incident on the SNSPD to maximize the detection efficiency. A small fraction of the reflected light is sent to an InGaAs photodiode, which is used to measure the cavity reflection spectrum while tuning the cavity resonance.

## 3.2 Tuning the cavity resonance

The resonance of a PhC is tuned in the cryostat by manipulating a thin layer of $N_2$ ice on the cavity [10, 11]. Clean, dry $N_2$ is applied through a nozzle to form an ice layer, which can decrease the cavity resonance frequency by up to 1 THz. To increase the resonance frequency, we desorb the ice using either laser light ($\sim 100\,\mu$W of 1536 nm light resonant with the cavity), or by heating the cryostat cold finger with a resistive heater. These techniques allow us to fix the resonance frequency with a precision better than 100 MHz. Condensation of background gas in the cryostat causes a persistent redshift at a rate of $\sim$1 GHz/hr.

## 3.3 Photon collection efficiency

The collection efficiency $\eta$ of light from the PhC can be determined from several independent measurements. The contributions are inefficiencies in the cavity ($\eta_{\text{cav}} = 0.16$), at the fiber-waveguide interface ($\eta_{\text{wg}} = 0.46$), in the optical fiber network between the cryostat and the SNSPD ($\eta_{\text{fib}} = 0.8$), and the finite detection efficiency of the SNSPD ($\eta_{\text{det}} = 0.67$). The product of these, $\eta = 0.04$, represents the probability for a photon in the cavity to result in a click on the detector.

The fiber-waveguide interface losses are determined by measuring the power reflection coefficient of the PhC away from the cavity resonance, and taking the square root to convert round-trip losses into one-way losses. The cavity losses are determined from the ratio of the on- and off-resonance reflection level ($C = 0.46$, Fig. 1e of the main text) of the cavity, which depends on the ratio of the internal ($\kappa_{\text{int}}$) and waveguide ($\kappa_{\text{wg}}$) loss channels from the cavity as $C = (1 - 2\eta_{\text{cav}})^2$, where $\eta_{\text{cav}} = \kappa_{\text{wg}}/(\kappa_{\text{wg}} + \kappa_{\text{int}})$. While this measurement cannot distinguish which of the two $\kappa$ terms is greater, a separate phase-sensitive measurement of the cavity reflection using a fast modulator and photodiode together with a vector network analyzer determined that $\kappa_{\text{wg}} < \kappa_{\text{int}}$. The detector efficiency is measured using a calibrated power meter and neutral density filters, and the fiber network transmission is directly measured.

The actual collection efficiency of light from an Er ion is determined by the product of $\eta$ and the probability for an ion in the excited state to decay by emitting a photon into the cavity. Given the total decay rate $\Gamma = P\Gamma_0 + \Gamma_0$, where the first term represents the cavity decay and the second term the bare decay rate, the probability to decay into the cavity is given by $\eta_{\text{Er}} = P/(P+1)$. For the Purcell factors presented here ($P > 100$), $\eta_{\text{Er}} \approx 1$.

## 3.4 Ions measured

The measurements in Fig. 3 of the main text were performed using a single ion ("A") with a transition frequency of 6.513 GHz, in the units of Fig. 2. The measurement in Fig. 4 was performed using the ion ("B") at 520 MHz, shown in Fig. 2b, to avoid overlapping transitions with nearby ions in large magnetic fields. The maximum Purcell factor for ion B is smaller than that for ion A, with a maximum value around 130. Ion B was also used for the measurements in Figs. S6, S7 and ??.

## 3.5 Minimum value of $g^{(2)}(0)$ in the presence of noise

Accidental coincidences from background light and dark counts limit the value of $g^{(2)}(0)$. This limit can be computed by representing the photons detected in a given bin $i$ ($n_i$) as the sum of two contributions: photons from a single atom ($n_i^{\text{ion}}$), and uncorrelated background ($n_i^{\text{bg}}$). Then:

$$g^{(2)}(0) = \frac{\langle n_i^2 \rangle}{\langle n_i \rangle^2} = \frac{\langle (n_i^{\text{ion}})^2 \rangle + \langle (n_i^{\text{bg}})^2 \rangle + 2\langle n_i^{\text{ion}} \rangle \langle n_i^{\text{bg}} \rangle}{\langle n_i^{\text{ion}} + n_i^{\text{bg}} \rangle^2} = \frac{2A+1}{(A+1)^2}$$

where $A = \langle n_i^{\text{ion}} \rangle / \langle n_i^{\text{bg}} \rangle$. We estimated the background count rate for the data in Fig. 3c of the main text by applying the same excitation sequence with the laser detuned by 30 MHz from the ion transition. From this measurement, we obtain $A = 5.5 \pm 1.6$, which puts a lower bound on $g^{(2)}(0) > 0.29 \pm 0.06$. The background count rate has contributions from both dark counts (35%) and weakly coupled ions (65%). Note that this value of $A$ is lower than its apparent value in Fig. 3a, because the $g^2$ measurement was performed with a higher duty cycle excitation (10 μs pulse with a 100 μs repetition period, compared to 400 μs period in Fig. 3a of the main text), which leads to more background ion excitation.

## 4 Temperature and magnetic field dependence of the spin-lattice relaxation time, $T_1$

The longest spin relaxation time measured for the $Z_1$ ground state in Er:YSO is 4.2 s [13], at a temperature of 20 mK and a magnetic field of 0.27 T. However, modeling the field- and temperature-dependence of the specific spin-lattice relaxation mechanisms involved leads to a prediction that dramatically longer relaxation times, exceeding $10^3$ seconds, should be achievable at higher temperatures ($\approx 1$ K) but lower magnetic fields. Figure S5 shows the expected behavior at several magnetic fields, together with several measurements from the literature.

The relaxation time $T_1$ of the ground state spin is limited by phonon-mediated processes. At cryogenic temperatures, direct, Raman and Orbach processes contribute to the total rate [14], with the functional form:

$$T_1^{-1} = A_d \left( \frac{g\mu_B B}{h} \right)^5 \coth\left( \frac{g\mu_B B}{2kT} \right) + A_r T^9 + A_o e^{-\Delta/(kT)} \tag{1}$$

Here, $g\mu_B B$ is energy difference between the two $Z_1$ spin sublevels in the magnetic field $B$, $k$ is Boltzmann's constant, and $h$ is Planck's constant. The coefficients describing the direct, Raman, and

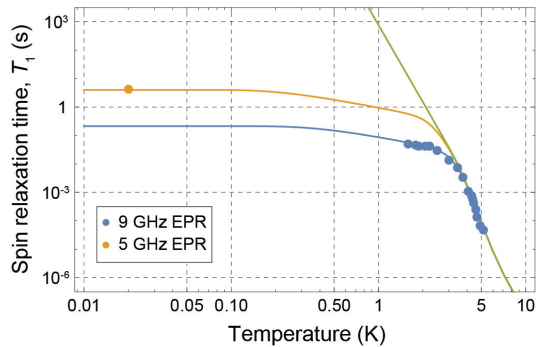

**Figure S5 | Temperature dependence of the spin $T_1$.** The blue points are data from Ref. [12] (X-band EPR, $g\mu_B B/h \approx 9$ GHz, site 1), while the yellow point is from Ref. [13] ($g\mu_B B/h \approx 5$ GHz). The blue line represents the model from Ref. [12] (Eq. 1), while the yellow line shows the same model with $g\mu_B B/h$ changed from 9 GHz to 5 GHz. The green line shows the same model with $g\mu_B B/h = 0.1$ GHz, illustrating the potential for extremely long spin-lattice relaxation times in the presence of small magnetic fields.

Orbach processes have previously been measured for Er:YSO [9, 12] using bulk, X-band ESR to be $(A_d, A_r, A_o) = (5.0 \times 10^{-5} \text{ s}^{-1}\text{GHz}^{-5}, 1.3 \times 10^{-3} \text{ s}^{-1}\text{K}^{-9}, 2.5 \times 10^{10} \text{ s}^{-1})$, with an Orbach intermediate state energy of $\Delta = 6.4$ meV.

The temperature dependence of the Raman and Orbach processes is very steep, restricting $T_1$ to values less than 10 ms for $T > 3$ K. In large magnetic fields, where $g\mu_B B/h$ is more than a few GHz, the direct process sets in around 2–3 K, preventing dramatic improvements at lower temperatures. However, the direct process is highly suppressed in low magnetic fields, as $B^4$, which should allow $T_1 > 10^3$ s, if $g\mu_B B/h < 100$ MHz. To the best of our knowledge, this regime has not been experimentally explored, as most studies of spin-lattice relaxation in Er:YSO are performed with electron paramagnetic resonance (EPR), which relies on large magnetic fields for spin polarization. What other limits arise in this regime (for example, from coupling to nuclear spins or other paramagnetic impurities) is unclear.

## 5  Saturation behavior of a single ion

We have repeated the single-ion PLE spectroscopy (Fig. 2) over a wide range of input powers to probe the saturation of the photon emission rate. At all powers, the spectrum qualitatively resembles a Gaussian peak on a flat background (Fig. S6a). While the background level increases linearly with input power, the height of the peak corresponding to a single ion transition saturates near 0.02 detected photons per excitation pulse (Fig. S6b). After accounting for the independently measured detection efficiency of 0.04 (Sec. 3.3), this number is consistent with 0.5 emitted photons per excitation pulse, as expected for saturated excitation in the presence of dephasing. The linear background likely originates from a large number of weakly coupled ions, which do not saturate and have a flat distribution in frequency.

To understand the saturation behavior of the ion, we model it as a two-level system and numerically solve the optical Bloch equations. The model parameters are all independently measured. The Rabi frequency is $\Omega = \sqrt{N_{\text{ph}}}\, g$, where $g = 2\pi \times 1.3$ MHz is the single-photon Rabi frequency determined from the Purcell factor (130 for this ion), and $N_{\mathrm{ph}}$ is the average intracavity photon number, related to the input power $P_{\mathrm{in}}$ as $N_{\mathrm{ph}} = 4\eta_{\mathrm{cav}} \frac{P_{\mathrm{in}}/\hbar\omega}{\kappa}$ (Sec. 3.3). The excited state decay rate is $\Gamma = 2\pi \times 1.8$ kHz. We also incorporate an additional excited-state dephasing rate $2\Gamma_{\mathrm{d}} = 2\pi \times 6.2$ MHz, chosen to match the observed linewidths at low power (Fig. S6c). The duration of the excitation pulse is 10 $\mu$s. The results of this model are in good agreement with the observed power-dependence of the fluorescence amplitude, as shown in Fig. S6b. The linewidth agrees qualitatively, but not quantitatively, with the experimental observations.

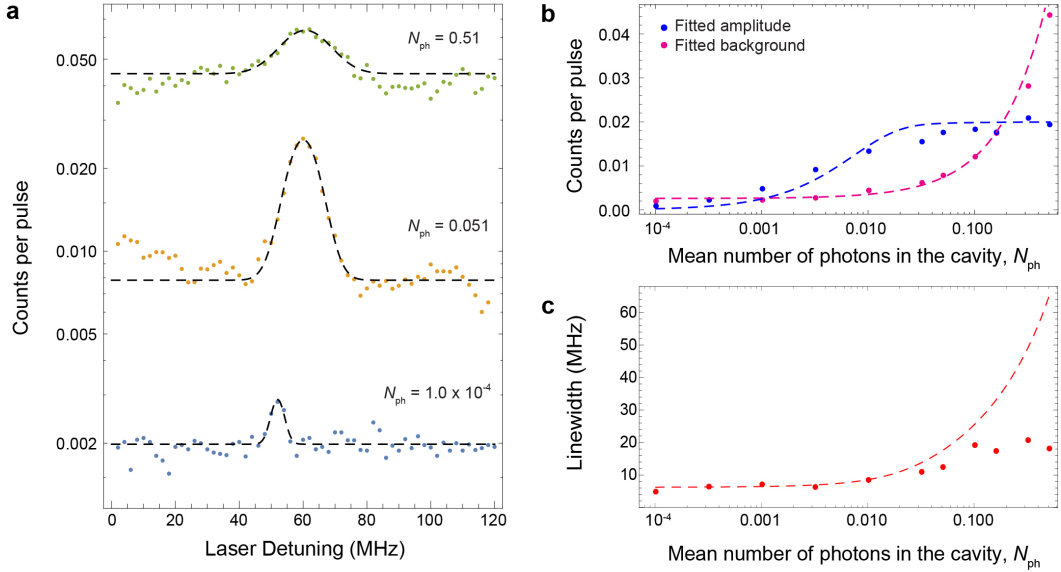

**Figure S6 | Saturation spectra and power broadening of a single $\text{Er}^{3+}$ ion. a,** PLE spectra recorded at several pump powers. The excitation power is indicated in units of the mean intracavity photon number $N_{\text{ph}}$. Each spectrum is fit to a Gaussian with a flat offset (black dashed line). **b,** Peak heights (blue) and offsets (magenta) determined from the Gaussian fits, for various excitation powers. The offsets increase linearly with power (dashed line), while the amplitudes are well-described by a numerical model (see Sec. 5). **c,** The linewidths (FWHM) from the Gaussian fits, for various excitation powers. The dashed line is from the same model as (b).

# 6 Additional data

In this section, we present several additional measurements, with discussions in the captions below the figures.

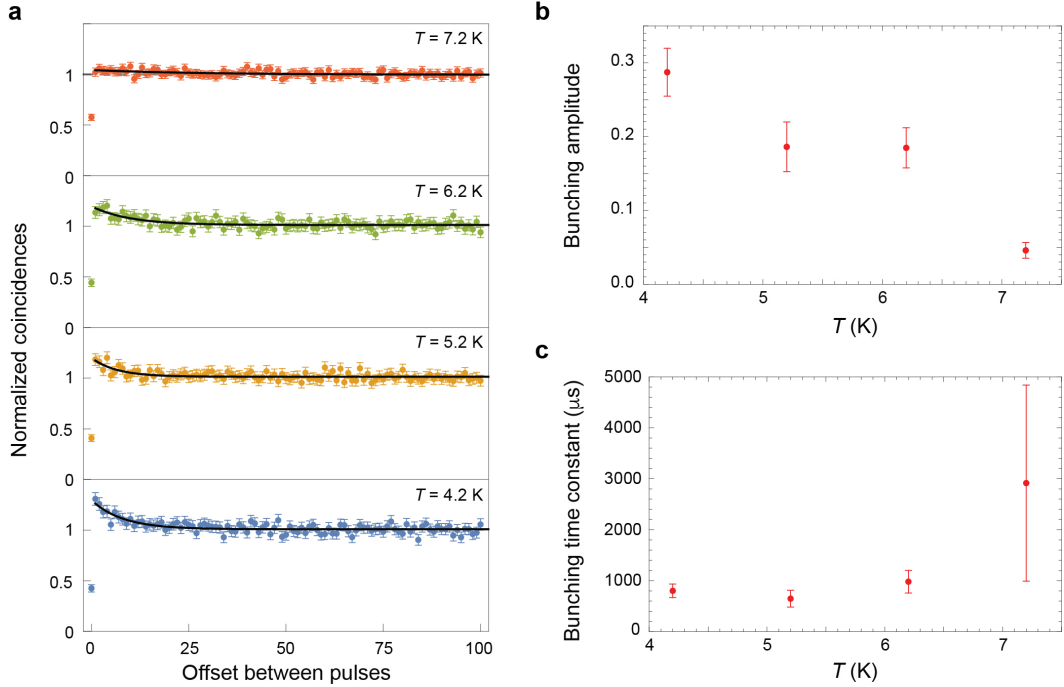

**Figure S7 | Temperature dependence of bunching in $g^{(2)}$. a**, Second-order autocorrelation function, $g^{(2)}$, of the fluorescence from a single ion at the indicated cryostat temperatures. In all curves, a large excitation power was used to reduce the averaging time necessary to acquire the bunching signal, at the expense of additional background that increases $g^{(2)}(0)$ slightly. The black curves are exponential fits to the bunching feature in $g^{(2)}$. **b-c**, Temperature dependence of the fitted bunching amplitudes (b) and time constants (c), where the conversion between pulse offset and time is performed using the pulse repetition period of 100 $\mu$s. The absence of a sharp temperature dependence in the time constant suggests that the origin of the bunching is not spin-lattice relaxation of the ground state spin (Sec. 4). This is additionally supported by independent measurements showing that the bunching behavior is the same in both small and large magnetic fields, whether the spin transitions spectrally overlap or not (as in Fig. 4 in the main text).

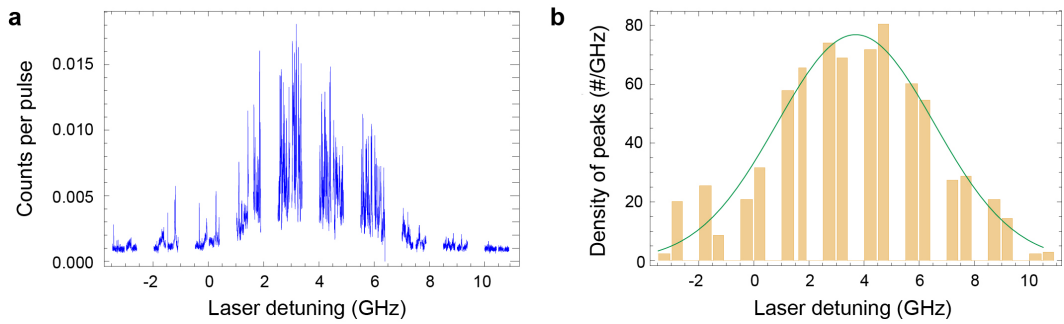

**Figure S8 | Counting peaks in the $Er^{3+}$ spectrum. a,** PLE data from Fig. 2a in the main text. **b,** A plot of the density of Lorentzian peaks that could be fit to the data. The Lorentzian widths are fixed, and only peaks with an amplitude $3\sigma$ above dark counts are included. The fitted Lorentzians overlap significantly near the center of the inhomogeneous distribution, so the counting results are not exact. The resulting spectral density of peaks is approximately Gaussian with a standard deviation of ($\sigma = 2.9$ GHz). We estimate that at least 500 lines are visible, using the Gaussian fit to interpolate number of peaks the missing spectral regions.

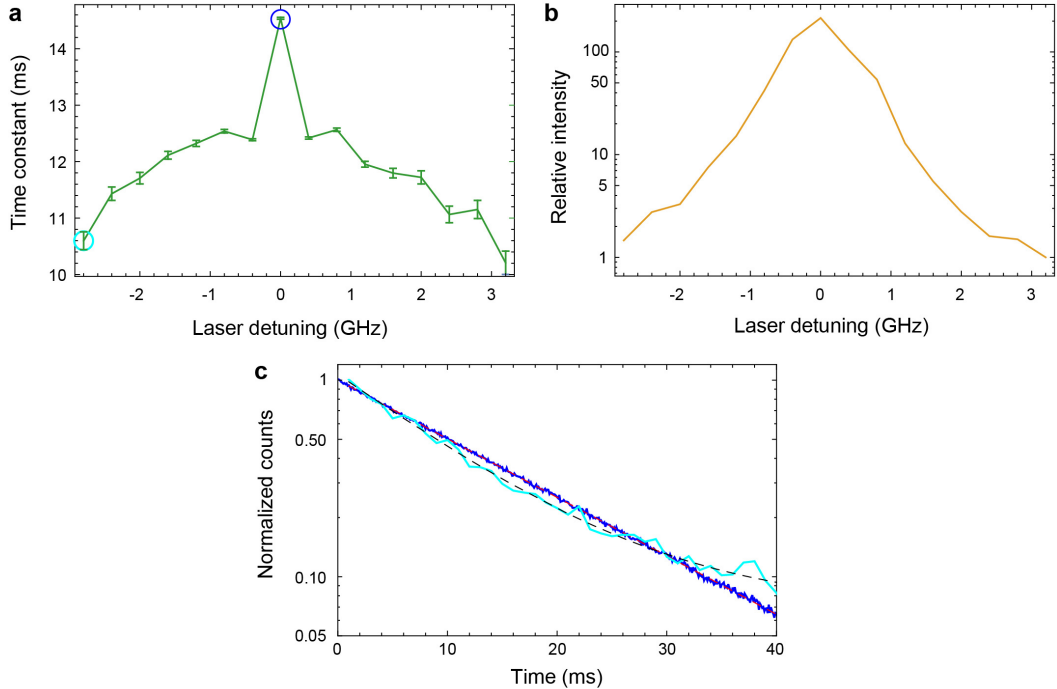

**Figure S9 | Measurement of the bulk Er:YSO excited state decay rate. a,** Fluorescence lifetime of the $^4I_{13/2}$ $Y_1$ state (site 1) in a 50 ppm Er:YSO crystal, as a function of the detuning of the excitation laser from the center of the inhomogeneous distribution. **b,** The associated fluorescence intensity for each of the detunings plotted in panel (a). **c,** Example time traces for detunings of -2.8 GHz (cyan) and 0 GHz (blue). Previous work has shown that the fluorescence lifetime can increase as a result of radiation trapping in dense samples [7, 15]. In panel (a), we find that the measured lifetime is largest at the detuning where the optical depth is highest, and decreases as the laser is detuned. The increase in the excited state lifetime at zero detuning may be associated with radiation trapping in the crystal, as the optical depth of the crystal is approximately 1 at this frequency. However, the decrease at large detunings beyond 1 GHz cannot plausibly be attributed to radiation trapping, because the optical depth decreases much more rapidly than the lifetime (using the fluorescence amplitude in (b) as a proxy for optical depth). Instead, this may result from strain or other effects that alter the crystal field levels and shift the transition away from the center of the inhomogeneous distribution. To avoid these complications, we rely on the previously reported literature value of $\tau_0 = 11.4$ ms to compute $\Gamma_0$, which was measured at the center of the inhomogeneous distribution in a lower density sample in Ref. [9].